\documentclass[prl,onecolumn,amsmath]{revtex4}
\def\moyal{\overleftrightarrow{\nabla}}
\begin{document}
\title{Stochasticity, decoherence and an arrow of time from the 
discretization of time?}
\author{M. C. Valsakumar}
\affiliation{Materials Science Division, Indira Gandhi Centre 
for Atomic Research, Kalpakkam 603102, India\\
E-mail: valsa@igcar.ernet.in}
\begin{abstract}{Certain intriguing consequences of the 
discreteness of time on 
the time evolution of dynamical systems are discussed. 
In the discrete-time classical mechanics proposed here, 
there is an {\it arrow of time} 
that follows from the 
fact that the replacement of the time derivative by the 
backward difference operator 
alone can preserve the non-negativity of the phase space density. 
It is seen 
that, even for free particles, all the degrees of freedom 
are {\it correlated} in 
principle. The forward evolution of 
functions of phase space variables by a finite 
number of time steps, in this discrete-time mechanics, 
depends on the entire 
continuous-time history in the interval $[0, \infty]$. 
In this sense, discrete 
time evolution is {\it nonlocal} in time from a 
continuous-time point of view. A  
corresponding quantum mechanical treatment is possible {\it via} the density 
matrix approach. The interference between 
non-degenerate quantum mechanical states 
decays exponentially. This {\it decoherence} 
is present, in principle, for all systems; 
however, it is of practical importance 
only in macroscopic systems, or in 
processes involving large energy changes.}
\end{abstract}
\maketitle

\section{Introduction}
{\it Time} is an enigma, and many philosophers and scientists have tried to 
ponder over its true meaning. However, most physicists and mathematicians 
consider the concepts of space and time to be intuitively obvious, and view 
space-time as an inert and infinitely divisible continuum in which `events' 
unfold. The advent of the general theory of relativity led to the insight that 
space-time is dynamical and that the gravitational field should be identified 
with certain properties of the space-time continuum. 
Subsequent efforts at synthesizing gravity and quantum 
mechanics, either with a background space as in string theory 
\cite{Danielsson2001}, or without a background space 
as in loop quantum gravity 
\cite{Ashtekar2001,Rovelli2003}, strongly suggest that 
space-time at the most fundamental level has a granular nature. 
In loop quantum gravity, the spatial Riemannian geometry is discrete 
\cite{Bojowald2000} with the volume of space 
quantized in units of $l_p^3$, where the Planck length     
$l_p = (\hbar G/c^3)^{1/2} \approx 1.6\times 10^{-35}$ m. 
Further, the time evolution takes place in discrete time 
steps \cite{Bojowald2001a} of the order of the Planck time $t_p =  
(\hbar G/c^5)^{1/2} \approx 5.4\times 10^{-44}$ s.

\medskip
There have been some attempts in the literature 
\cite{Schmidt,Jackson,Bondi,Yukawa,Yamamoto} to study the time evolution 
of Hamiltonian systems with time as a discrete parameter even before 
these exciting developments took place. 
In particular, Katayama and Yukawa 
argued that, just like matter, 
space-time also should have an indivisible element 
(an `elementary domain'). Yamamoto\cite{Yamamoto} also made an attempt 
to realize the elementary domain through quantum field theory in 
discrete time. In this paper, we propose a version of discrete-time mechanics 
as a precursor to the more complicated (and conceptually satisfying) discrete 
space-time mechanics. The idea is to explore the genuine consequences of 
discrete-time evolution, and not the
development of numerical methods to 
approximate continuous-time evolution with greater precision.   
In the equation of motion approach 
to classical mechanics (CM), Lee \cite{Lee1982,Lee1985} 
found that violation of 
time-translational invariance (that accompanies discretization of time) leads 
to non-conservation of energy. He therefore developed a theory with time as 
a discrete dynamical variable. Subsequently, 
Jaroszkiewicz and coworkers 
\cite{Jaroszkiewicz1997} succeeded in developing an equation of motion 
approach to CM with time as a discrete parameter by invoking the discrete-time 
action principle of Cadzow\cite{Cadzow}. 

\medskip
We take a fresh look at the problem and adopt the 
phase space density approach. 
We start with the Liouville equation for the phase space density in classical 
mechanics, and propose its discrete-time analogue. 
The structure of this discrete-time 
Liouville equation ensures that {\it all} the constants of the motion in 
continuous-time mechanics are also constants of the motion in discrete-time 
mechanics. The basic premises that underlie 
the derivation of this equation and 
the consequences of the discreteness of time (the 
emergence of stochasticity, correlation and 
non-locality) are discussed in Section 2. 
The mathematical details are relegated 
to Appendices A and B. We then quantize the discrete-time Liouville equation 
to obtain the time evolution of the Wigner distribution function 
(equivalently, 
the density matrix). It is seen that the 
discretization of time leads to energy 
super-selection and decoherence. 
These points are discussed in Section 3. 
Even though the present formalism is motivated by 
Hamiltonian mechanics, we have 
attempted to study the consequences of discrete time on more general forms of 
dynamical systems. In particular, the effect of the 
discreteness of time on the 
sensitive dependence on initial conditions of (chaotic) nonlinear dynamical 
systems is discussed briefly in Section 4. Finally, 
some conclusions are drawn 
in Section 5. Before we proceed, 
we would like to emphasize that the 
fundamental unit of time $\tau$ that appears 
in our formalism need {\it not} 
be equal to the Planck time. We also remark that 
the usual continuous-time mechanics is 
recovered in the limit $\tau \to 0$.

\section{Classical mechanics in discrete time} 
We consider a dynamical system with $l$ degrees of freedom described by the 
Hamiltonian $H(\vec{x}\,,\,\vec{p})$, 
where $x_i$ and $p_i$ are the coordinates and 
canonical momenta, respectively. 
The time evolution of the phase space density 
$\rho_{\rm ct}(\vec{x}\,,\,\vec{p}\,,\,t)$ is given by the Liouville equation 
\begin{equation}
\frac{\partial}{\partial t}\rho_{\rm ct}(\vec{x}\,,\,\vec{p}\,,\,t) = 
\lbrace{H\,,\,\rho_{\rm ct}}\rbrace = L\rho_{\rm ct}(\vec{x}\,,\,\vec{p}\,,\,t)
\label{eqliouvillec}
\end{equation}
where $\lbrace{H\,,\,\rho_{\rm ct}}\rbrace$ is the Poisson bracket of $H$ with 
$\rho_{\rm ct}(\vec{x}\,,\,\vec{p}\,,\,t)$, 
and $L$ is the Liouville operator
\begin{equation}
\nonumber
L = \sum_{i=1}^{l}\left(\frac{\partial H}{\partial x_i} 
\frac{\partial}{\partial p_i} - \frac{\partial H}{\partial p_i} 
\frac{\partial}{\partial x_i}\right).
\end{equation}
Throughout this paper, the suffixes `ct'  and `dt' 
as in $F_{\rm ct}$ and $F_{\rm dt}$ shall 
denote values of the phase space function 
$F$ in continuous-time mechanics and discrete-time mechanics, respectively.
The formal solution of Eq. (\ref{eqliouvillec}) is given by
\begin{equation}
\rho_{\rm ct}(\vec{x}\,,\,\vec{p}\,,\,t) = e^{tL}\, 
\rho_{\rm ct}(\vec{x}\,,\,\vec{p}\,,\,0)
\label{eqformal}
\end{equation}
where $\rho_{\rm ct}(\vec{x}\,,\,\vec{p}\,,\,0)$ 
is the initial phase space density.  
Throughout this paper, we consider the deterministic initial condition
\begin{equation}
\rho_{\rm ct}(\vec{x}\,,\,\vec{p}\,, 0) =\delta(\vec{x} - \vec{x}(0))
\,\delta(\vec{p} - \vec{p}(0)).
\label{eqic}
\end{equation}
The value $F_{ct}(t)$ of the 
phase space function $f(\vec{x}, \vec{p})$ at 
time t is given by
\begin{equation}
\nonumber
F_{ct}(t) = \left<f(\vec{x}, \vec{p})\right>_{\rm ct} 
= \int \!d\vec{x} \int \!d\vec{p}\, 
f(\vec{x}, \vec{p}) \,\rho_{\rm ct}(\vec{x}\,,\,\vec{p}\,,\,t).
\end{equation}
Our aim is to ``deduce'' 
the discrete-time analogue of Eq. (\ref{eqliouvillec}). 
Since there is no unique definition of a discrete time derivative, we do this 
by first stipulating the conditions to be satisfied by the equation that 
governs the time evolution of the phase space density 
$\rho_{\rm dt}(\vec{x}\,,\,\vec{p}\,,n)$ 
at discrete time $t$ = $n\tau$, where $\tau$ is the 
fundamental unit of time. Specifying the data 
at one instant of time is sufficient to obtain 
the solution of Eq. (\ref{eqliouvillec}). We demand that specifying 
$\rho_{\rm dt}(\vec{x}\,,\,\vec{p}\,,\,j)$  at some discrete time 
$t = j\tau$ should 
be sufficient to determine $\rho_{\rm dt}(\vec{x}\,,\,\vec{p}\,,\,j')$ 
at any other  discrete 
time $t' = j'\tau$. This implies that the equation to be discovered should 
essentially be a first-order difference equation in time. 
We thus obtain the 
most general form of the discrete-time Liouville equation 
for forward evolution in the form   
\begin{equation}
\frac{\rho_{\rm dt}(\vec{x}\,,\,\vec{p}\,,\,n+1) 
- \rho_{\rm dt}(\vec{x}\,,\,\vec{p}\,,\,n)}{\tau}
= L\left[\alpha \,\rho_{\rm dt}(\vec{x}\,,\,\vec{p}\,,\,n) + 
\beta \,\rho_{\rm dt}(\vec{x}\,,\,\vec{p}\,,\,n+1)\right],
\label{eqliouvilled}
\end{equation}
where $0 \le \alpha\,,\,\beta \le 1$
and $\alpha + \beta = 1$. Special cases 
of Eq. (\ref{eqliouvilled}) are the forward and backward difference schemes  
obtained from Eq. (\ref{eqliouvilled}) by setting $\alpha = 1 $ and
$\alpha = 0$,  
respectively. Restrictions on the possible values of $\alpha$ can be obtained 
by examining the time evolution operator $T_n$ defined through the equation
\begin{equation}
\rho_{\rm dt}(\vec{x}\,,\,\vec{p}\,,\,n) = 
\left(\frac{1 + \alpha\tau L}{ 1 - \beta\tau L}\right)^n 
\rho_{\rm dt}(\vec{x}\,,\,\vec{p}\,,\,0) = T_n\, 
\rho_{\rm dt}(\vec{x}\,,\,\vec{p}\,,\,0).
\label{eqtimeevol}
\end{equation}
We assume that the Hamiltonian and the 
manifold in which the dynamics takes 
place are such that $L$ is sufficiently well defined. $L$ is then a 
skew-Hermitian operator, and hence its spectrum is pure imaginary. 
Using the relation
\begin{equation}
\nonumber
\int{f(\vec{\Omega})\left[G(L) g(\vec{\Omega})\right] d\vec{\Omega}} = 
 \int{\left[G(-L) f(\vec{\Omega})\right] g(\vec{\Omega}) d\vec{\Omega}}
\end{equation} 
where $\vec{\Omega}$ = $(\vec{x}\,,\, \vec{p})$, we can show that every 
quantity that is conserved in the continuous time context is also 
conserved under the time evolution scheme proposed here. 

\medskip
It is clear that it is only when $\alpha = \frac{1}{2}$ that 
${T_n}$ is unitary and the evolution given by Eq. 
(\ref{eqtimeevol}) is consistent with time reversal invariance. 
$T_n$ is an unbounded operator (a bounded function 
would evolve to an unbounded one under the action of $T_n$) in 
the limit $n\rightarrow \infty$ for $\alpha > \frac{1}{2}$. 
Since this is undesirable, 
$\alpha$ must be less than $\frac{1}{2}$ for $n > 0$. 
Similarly, $T_n$ is unbounded 
in the limit $n \rightarrow -\infty$ for $\alpha < \frac{1}{2}$. 
Hence $\alpha$ 
must be greater than $\frac{1}{2}$ 
for $n < 0$. Thus, an arrow of time emerges naturally 
if $\alpha \ne \frac{1}{2}$.

\medskip
A more stringent constraint on the value of $\alpha$ can be obtained by 
demanding that the phase space density $\rho_{\rm dt}(\vec{x}\,,\,\vec{p}\,, n)$ be 
{\it non-negative}. We examine this problem in two different ways, and show 
that zero is the only permissible value of $\alpha$. 

\subsection{\,\,Method 1:}
Let us make a formal transformation 
\begin{equation}
\nonumber
(x_1, ...., x_l; p_1, ...., p_l) \rightarrow (\xi_1, ...., \xi_l; 
\eta_1, ......, \eta_l), 
\end{equation}
such that $L = - \partial/\partial \xi_1\,$. This transformation can 
be worked out explicitly for systems with quadratic
Hamiltonians. While  
global realization of such a transformation for systems with non-quadratic 
Hamiltonians appears to be doubtful, the analysis is instructive,  
and we proceed with it. For the deterministic initial condition 
Eq. (\ref{eqic}), the corresponding initial condition in the transformed 
variables is given by
\begin{equation}
\nonumber
\bar{\rho}_{\rm dt}(\vec{\xi},\vec{\eta}, 0) = 
\prod_{i=1}^{l}\delta\big(\xi_i - \xi_i(0)\big)\,
 \delta\big(\eta_i - \eta_i(0)\big).  
\end{equation}
The time evolution of 
$\bar{\rho}_{\rm dt}(\vec{\xi},\vec{\eta}, n)$ is given by 
Eq. (\ref{eqtimeevol}) to be
\begin{equation}
\nonumber
\bar{\rho}_{\rm dt}(\vec{\xi},\vec{\eta}, n) = 
g_n(\xi_1) \,\delta\big(\eta_1 - \eta_1(0)\big) 
\prod_{i=2}^{l}\delta\big(\xi_i - \xi_i(0)\big) \,\delta
\big(\eta_i - \eta_i(0)\big),
\end{equation}
where 
\begin{equation}
\nonumber
g_n(\xi_1) = \left(\frac{1 - \alpha\tau \,\partial/\partial \xi_1}
{1 + \beta\tau \,\partial/\partial \xi_1}\right)^n 
\delta(\xi_1 - \xi_1(0)).
\end{equation}
As shown in Appendix A, we find that zero is the only allowed 
value of $\alpha$ which renders $\bar{\rho}_{\rm dt}(\vec{\xi},\vec{\eta}, n)$ 
non-negative and $T_n$ bounded. For this case, 
a simple expression can be derived 
for $g_n(\xi_1)$, namely,
\begin{equation}
g_n(\xi_1) = \begin{cases}
\frac{1}{(n-1)!\tau} \left(\frac{\xi_1 -\xi_1(0)}{\tau}\right)^{n-1} 
\exp{\left(-\frac{\xi_1 - \xi_1(0)}{\tau}\right)}, 
\,\, {\rm for}\,\, \xi_1 > \xi_1(0) \cr
\ 0, \quad\quad\quad\quad\quad\quad\quad\quad\quad\quad\quad\quad
\quad\quad\quad\quad {\rm for}\,\, \xi_1 < \xi_1(0).\end{cases}
\label{eqgnxi1final}
\end{equation}
We thus see that the backward difference scheme is the only acceptable 
generalization of the time derivative to the discrete domain. 

\subsection{\,\,Method 2:}
Here we obtain a relation between $\rho_{\rm ct}(\vec{x}\,,\,\vec{p}\,,\,t)$ and 
$\rho_{\rm dt}(\vec{x}\,,\,\vec{p}\,,\,n)$ by resorting to the generating function 
technique. As shown in Appendix B, we  arrive at the same conclusion:
namely, that 
zero is the only permissible value of $\alpha$ that renders 
$\rho_{\rm dt}(\vec{x}\,,\,\vec{p}\,,\,n)$ non-negative. This analysis leads to the  
following intriguing relation for $n > 0$:
\begin{equation}
\rho_{\rm dt}(\vec{x}\,,\,\vec{p}\,,\,n) = \frac{1}{(n-1)!} 
\int_{0}^{\infty} du \ e^{-u} u^{n-1} \rho_{\rm ct}(\vec{x}\,,\,\vec{p}\,,\,\tau u).
\label{eqnonlocal}
\end{equation}
The average value $F_{\rm dt}(n)$ of any 
function $f(\vec{x}, \vec{p})$ at time $t=n\tau$ is 
given by 
\begin{equation}
F_{dt}(n) = \left<f(\vec{x}, \vec{p})\right>_{dt,n} = \frac{1}{(n-1)!} 
\int_{0}^{\infty} du \ e^{-u} u^{n-1} F_{ct}(\tau u),
\label{eqnonlocal2}
\end{equation}
where $F_{ct}(\tau u)$ is the phase space 
average in continuous-time mechanics at 
time $t$= $\tau u$. Such quantities are {\it nonlocal} in time 
from the continuous-time 
point of view! The discreteness of space-time thus leads 
to non-locality in time in this sense.

\medskip
Both Eqs. (\ref{eqgnxi1final}) and (\ref{eqnonlocal}) imply that an element of 
stochasticity appears in 
the discrete-time classical mechanics described in this 
paper. Equation (\ref{eqgnxi1final}) suggests that the stochasticity is 
essentially in one variable, $\xi_1\,$, 
and that its conditional probability density is 
a gamma distribution for all Hamiltonian systems. 
Equation (\ref{eqnonlocal}) 
implies a random walk in time. 
Thus $\xi_1$ can be identified with a kind of ``internal time''.

\subsection{\,\,Examples}
We now illustrate the peculiarities of discrete-time evolution with the 
help of two examples: $(i)$ a collection of free particles, and $(ii)$
a collection of  harmonic oscillators.\\

\noindent
$(i)$ \,{\it Free particles}:\\

\noindent
For a collection of free particles described by the Hamiltonian $H$ = 
$\sum_{i=1}^{l} p_i^2/(2m_i)$ with deterministic initial conditions, 
we get the following results for the averages 
of $x_i\,,\,p_i\,,\,x_i^2$ and $p_i^2$ 
at discrete time $n\tau$:
\begin{equation}
\nonumber
\left<x_i\right>_{{\rm dt},n} = x_i(0) + \frac{p_i(0)n\tau}{m_i}, 
\quad \left<p_i\right>_{{\rm dt},n} = p_i(0), 
\end{equation}
\begin{equation}
\nonumber
\left<x_i^2\right>_{{\rm dt},n} 
= \left(x_i(0) + \frac{p_i(0)n\tau}{m_i}\right)^2 + 
\frac{np_i^2(0)\tau^2}{m_i^2}, \quad  
\left<p_i^2\right>_{{\rm dt},n} = p_i^2(0). 
\end{equation}
Since $\left<p_i^2\right>_{{\rm dt},n}$ = $p_i^2(0)$, 
$H$ is conserved. However, 
\begin{equation}
\nonumber
\left<x_i^2\right>_{{\rm dt},n} - \left<x_i\right>_{{\rm dt},n}^2 = 
n\frac{p_i^2(0)\tau^2}{m_i^2} = 
D_i t
\end{equation}
where $D_i = p_i^2(0)\tau/m_i^2$. Hence the motion 
of the particles is {\it diffusive}. A similar calculation  
shows that 
\begin{equation}
\nonumber
\left<x_ix_j\right>_{{\rm dt},n} -
\left<x_i\right>_{{\rm dt},n}\left<x_j\right>_{{\rm dt},n} = 
n\frac{p_i(0)p_j(0)\tau^2}{m_im_j},
\end{equation}
which implies that even the motion of {\it non}-interacting 
particles is correlated in the foregoing  
sense. \\

\noindent
$(ii)$\,{\it Harmonic oscillators}:\\

\noindent
Next, consider  a collection of harmonic oscillators described by the 
Hamiltonian $H = \sum_{i=1}^{l}\frac{1}{2}(p_i^2 
+ x_i^2)$, where we have taken all the masses and 
frequencies to be identical and set
$m_i= 1,\, \omega_i=1$, for simplicity. The analysis 
of this case is straightforward. The final expressions 
for the first and second moments of $x_i$ and $p_i$ at time $n\tau$ 
are
\begin{equation}
\nonumber
\left<x_i\right>_{{\rm dt},n} 
= \frac{r_i(0)}{(1+\tau^2)^{n/2}}\sin\,\big(n\phi+\theta_i(0)\big), 
\end{equation}
\begin{equation}
\nonumber
\left<p_i\right>_{{\rm dt},n} 
= \frac{r_i(0)}{(1+\tau^2)^{n/2}}\cos\,\big(n\phi+\theta_i(0)\big), 
\end{equation}
\begin{equation}
\nonumber
\left<x_i^2\right>_{{\rm dt},n} 
= \frac{r_i^2(0)}{2}\left[
1 + \frac{\cos\,\big(n\phi'+2\theta_i(0)\big)}
{(1+4\tau^2)^{n/2}}\right],
\end{equation}
\begin{equation}
\nonumber
\left<p_i^2\right>_{{\rm dt},n} 
= \frac{r_i^2(0)}{2}\left[1 - 
\frac{\cos\,\big(n\phi'+2\theta_i(0)\big)}{(1+4\tau^2)^{n/2}}\right], 
\end{equation}
\begin{equation}
\nonumber
\left<x_ix_j\right>_{{\rm dt},n} 
= \frac{r_i(0)r_j(0)}{2}\left[\cos\,\big(\theta_i(0)-\theta_j(0)\big) 
+ \frac{\cos\,\big(n\phi'+\theta_i(0)+\theta_j(0)\big)}
{(1+4\tau^2)^{n/2}}\right],
\end{equation}
where $\phi = \arctan\,\tau,\, 
\phi' = \arctan\,(2\tau), \,r_i^2 = x_i^2+p_i^2$, 
and $\theta_i=\arctan\,(x_i/p_i)$. We see that 
$\left<x_i^2\right>_{{\rm dt},n}$ 
- $\left<x_i\right>_{{\rm dt},n}^2$ 
and $\left<p_i^2\right>_{{\rm dt},n}$ - 
$\left<p_i\right>_{{\rm dt},n}^2$ are 
nonzero, signalling stochasticity in $x_i$ and $p_i$. 
However, the Hamiltonian $H$ is conserved. 
We also see that $\left<x_ix_j\right>_{{\rm dt},n}$ - 
$\left<x_i\right>_{{\rm dt},n}\left<x_j\right>_{{\rm dt},n}$ is
nonzero, implying that all the degrees 
of freedom are now correlated.

\section{Quantum mechanics in discrete time}
Quantum mechanics (QM) is a remarkably successful theory with no known 
physical phenomena that contradict it. Yet, it has no universally accepted 
interpretation and hence the adage, ``It is a theory that works for all 
practical purposes'' \cite{Bell}.  The founding fathers of QM have insisted 
that the results of {\it measurements} have to be expressed in classical 
terms. If the world is quantum mechanical, then classical mechanics (CM) 
should be contained within QM as a limiting case.  Now, a crucial ingredient 
of QM is the principle of superposition that follows from the linearity of 
the Hilbert space. Even though manifestations 
of quantum mechanical interference 
abound in the microscopic domain, the 
application of such a principle to the 
macro-world seems to lead to predictions that are counter-intuitive 
{\it vis-\`a-vis} our day-to-day experience. The Schr\"odinger 
cat paradox, invented by 
Schr\"odinger himself, 
clearly brings these issues into focus, and it suggests 
that, from among the multitude of superpositions 
allowed by the Schr\"odinger 
equation, only a few robust states are allowed for macroscopic systems. 
The so-called {\it decoherence program} 
\cite{Zurek1991,Giulini1996,Schlosshauer2003,Bacciagaluppi2003} shows how this 
may come about entirely within a quantum mechanical description, by invoking 
the unavoidable interaction of any given system with the external world. 
Entanglement between the system and the environment can cause super-selection, 
{\it i.e.}, the selection of a preferred set of states that are robust in 
spite of their immersion into the environment. Another consequence  
of the entanglement is environment-induced decoherence, which refers
to the  
suppression of interference between the preferred states chosen by the 
super-selection rule. In what follows, we show how energy super-selection and 
decoherence may arise from the discretization of time. 

\medskip
In continuous time, the density matrix $\hat{\rho}_{\rm ct}(t)$ satisfies the 
well known evolution equation 
\begin{equation}
\nonumber
\frac{\partial}{\partial t}\hat{\rho}_{\rm ct}(t) 
= \frac{1}{i\hbar}\big[\hat{H}, 
\hat{\rho}_{\rm ct}(t)\big],
\end{equation}
where $\hat{H}$ is the Hamiltonian operator and 
$[\hat{A}\,,\, \hat{B}]$ 
is the commutator of the operators $\hat{A}$ and $\hat{B}$. The usual 
quantization prescriptions whereby 
$\rho_{\rm ct}$ $\to$ $\hat{\rho}_{\rm ct}$ and the Poisson 
bracket $\lbrace{A, B}\rbrace \to [\hat{A}, 
\hat{B}]/(i\hbar)$, together with the procedure for going from continuous to 
discrete time (described in this paper), leads to the following evolution 
equation for the discrete-time density matrix $\hat{\rho}_{\rm dt}(n)$: 
\begin{equation}
\nonumber
\frac{\hat{\rho}_{\rm dt}(n+1) - \hat{\rho}_{\rm dt}(n)}{\tau} = 
\frac{1}{i\hbar} \left[\hat{H}\,,\, \hat{\rho}_{\rm dt}(n+1)\right] = 
\hat{L}\hat{\rho}_{\rm dt}(n+1)
\end{equation}
for $n$ $>$ $0$. This implies that
\begin{equation}
\nonumber
\hat{\rho}_{\rm dt}(n) = \left(1 - \tau\hat{L}\right)^{-n} \hat{\rho}_{\rm dt}(0).
\end{equation}
Let $\vert\alpha\rangle$ denote the eigenfunction of $\hat{H}$ 
with the eigenvalue 
$\epsilon_{\alpha}$, and $\hat{\rho}_{\rm dt}(0)$ the initial density matrix:
\begin{equation}
\nonumber
\hat{H} \vert\alpha\rangle = \epsilon_{\alpha}\vert\alpha\rangle, \ \ 
\hat{\rho}_{\rm dt}(0) = 
\sum_{\alpha,\beta}a_{\alpha,\beta}\,\vert\alpha\rangle\langle\beta\vert .
\end{equation}
We then get
\begin{equation}
\nonumber
\hat{\rho}_{\rm dt}(n) = \sum_{\alpha,\beta}a_{\alpha,\beta}\left[
1 + i\tau\left(\frac{\epsilon_{\alpha} - 
\epsilon_{\beta}}{\hbar}\right)\right]^{-n}
\vert\alpha\rangle\langle\beta\vert .
\end{equation}

It is clear that the diagonal elements of $\hat{\rho}_{\rm dt}(n)$ are 
time-invariant. The off-diagonal elements decay exponentially if the basis 
states are non-degenerate. They are, however, time invariant if the 
states $\vert\alpha\rangle$ and $\vert\beta\rangle$ are degenerate.  We thus 
find that, in principle, the interference involving states with different 
energies decays exponentially with a characteristic time $T_d$ given by
\begin{equation}
\nonumber
T_d = (2\tau)/\log{\left[1+(\Delta E\tau/\hbar)^2\right]},
\end{equation}
where $\Delta E$ is the difference in energy between the states. If $\tau$ 
is taken to be the Planck time ($5.4\times 10^{-44}\,{\rm s}$), we 
find the decay time to be greater than $10^{10}$ years if $\Delta E >
7 \,{\rm meV}$. 
Thus, a microscopic system prepared in a mixed state by superposing states 
separated in energy by a few meV, decoheres only in principle, and continues 
to be coherent in practice. 
However, for a macroscopic system with about $10^{20}$ 
particles, $T_d \sim 10^{-23}$ s for states 
with a $7 \,{\rm meV}$ change in the energy  per particle.

\medskip
The formalism described in this paper can be extended to quantum mechanical 
distribution functions (QDF) as well. In view of the non-uniqueness associated 
with the classical $\leftrightarrow$ quantum correspondence 
(i.e., the way to construct quantum 
mechanical operators corresponding to classical phase space functions), there 
are  many ways of defining QDFs \cite{LCohen}. We restrict ourselves to the 
Wigner distribution function, which follows from the Weyl correspondence rule. 
We can obtain a quantum mechanical description in 
discrete time by Wigner-Moyal 
quantization of the discrete-time Liouville equation. 
This is achieved by simply 
replacing the Poisson bracket by the Moyal bracket. The time evolution of the 
discrete-time Wigner distribution function $W_{\rm dt}(\vec{x}\,,\,\vec{p}\,, n)$ is 
given by the backward difference equation
\begin{equation}
\nonumber
\frac{W_{\rm dt}(\vec{x}\,,\,\vec{p}\,, n+1) 
- W_{\rm dt}(\vec{x}\,,\,\vec{p}\,, n)}{\tau} = 
H\left(\frac{2}{\hbar}\sin\left[\frac{\hbar}{2}\moyal\right]\right)
W_{\rm dt}(\vec{x}\,,\,\vec{p}\,, n+1),
\end{equation}
where $\moyal$ is the operator defined through the relation
\begin{equation}
\nonumber
A \moyal B = \lbrace{A,B}\rbrace .
\end{equation}
It may be noted that Eq. (\ref{eqnonlocal}) 
holds good in the context of quantum 
mechanics as well, with the replacements 
$\rho_{\rm ct}(\vec{x}\,,\,\vec{p}\,, t) \to  
\hat{\rho}_{\rm ct}(t) {\rm \ and \ } \rho_{\rm dt}(\vec{x}\,,\,\vec{p}\,, n) \to 
\hat{\rho}_{\rm dt}(n)$ for the density matrix, and $
\rho_{\rm ct}(\vec{x}\,,\,\vec{p}\,, t) \to 
W_{\rm ct}(\vec{x}\,,\,\vec{p}\,, t) {\rm \ and \ } 
\rho_{\rm dt}(\vec{x}\,,\,\vec{p}\,, n) 
\to W_{\rm dt}(\vec{x}\,,\,\vec{p}\,, n)$ for 
the Wigner distribution function.

\subsection{Connection with the Schr\"odinger equation}
It is interesting to ask if there exists a discrete-time 
Schr\"odinger equation whose solution is consistent with the 
time evolution of the density matrix described in this paper. If 
such an equation exists, then the eigenstate $\vert\alpha\rangle$ should 
evolve to $\vert\alpha\rangle_n = f(n,\alpha) \, \vert\alpha\rangle$, where 
$f(n,\alpha)$ is an unknown function that should satisfy the relation
\begin{equation}
f(n,\alpha)f^{\star}(n,\beta) = \left[1 + 
i\tau\left(\frac{\epsilon_{\alpha} - 
\epsilon_{\beta}}{\hbar}\right)\right]^{-n}.
\label{eqsch1}
\end{equation}
When $\alpha$ = $\beta$, Eq. (\ref{eqsch1}) 
reduces to $\vert f(n,\alpha)\vert^2 
= 1$, which implies that $fn,\alpha)$ is unimodular: thus 
$f(n,\alpha) = 
\exp\,[i \,\Theta(n,\alpha)]$, 
where $\Theta(n,\alpha)$ is a real number.  When this 
expression for $f(n,\alpha)$ is substituted in Eq. (\ref{eqsch1}), we get  
\begin{equation}
\nonumber
\Theta(n,\alpha) - \Theta(n,\beta) 
= n \arctan \left(\frac{\tau(\epsilon_{\alpha} 
- \epsilon_{\beta})}{\hbar}\right) +i \frac{n}{2} \log\left[1+ \left(
\frac{\tau(\epsilon_{\alpha} - \epsilon_{\beta})}{\hbar}\right)^2\right],
\end{equation} 
in which the left-hand side is real, whereas the right-hand side is 
complex unless $\tau$ = 0. Thus the functional equation (\ref{eqsch1})
does not have a solution, and hence we do not have a discrete-time 
Schr\"odinger equation consistent with the equation for the density matrix. 
This is logically consistent with the fact that the density matrix description 
predicts decoherence, 
which never can occur in the Schr\"odinger formalism for an 
isolated system.

\section{Nonlinear dynamics in discrete time}
We have so far concentrated on the 
classical and quantum mechanics of Hamiltonian 
systems. However, we do encounter more 
general dynamical systems while modeling numerous 
phenomena in physics, chemistry and biology. It would be of 
interest, therefore, 
to extend the prescription (for going from continuous time  
to discrete time evolution) outlined in this paper 
to dynamical systems described by the 
set of (in general, nonlinear) ordinary differential equations 
\begin{equation}
\nonumber
\frac{d}{dt}x_j(t) = f_j(\lbrace x_i\rbrace), \ \ j=1, \ ....,\ n.
\end{equation}
We have the equivalent phase space formulation
\begin{equation}
\nonumber
\frac{\partial}{\partial t}\rho_{\rm ct}(\lbrace x_i\rbrace, t) = 
L \rho_{\rm ct}(\lbrace x_i\rbrace, t) = 
-\sum_{j=1}^n\frac{\partial}{\partial x_j} \left[f_j(\lbrace x_i\rbrace) 
\rho_{\rm ct}(\lbrace x_i\rbrace, t)\right].
\end{equation}
We note that, presented in this form, 
there is no obvious distinction between 
linear and nonlinear equations of motion 
(the differences between them would be reflected 
in the spectral properties of the Liouville operator). As a matter of fact, 
finite-dimensional nonlinear differential 
equations of motion can be reformulated 
as linear differential equations in infinite dimensions by adopting
the Carleman embedding procedure\cite{Carleman} or
similar ones. 
However, from a practical point 
of view, there are significant differences between linear and
nonlinear problems. 
Of particular interest is the tendency of nonlinear systems to show chaotic 
behaviour --- 
{\it bounded and aperiodic evolution which shows sensitive dependence 
on initial conditions}. In view of 
the formula (see Eq. (\ref{eqnonlocal2}))  
that relates forward evolution in discrete time to the entire continuous-time 
history in the semi-infinite interval $[0, \infty]$, it is 
intuitively clear that sensitive dependence on initial conditions 
in discrete-time mechanics should be different from that of 
continuous-time mechanics. 
Some sort of ``rounding off'' of the instability 
should occur in the discrete 
time context, at least in some systems that show bounded evolution, as the 
following example would suggest. 

\medskip
Let the continuous-time solution for one of the 
dynamical variables, $x$, with 
initial value $a$, be given by the relation
\begin{equation}
x_{\rm ct}(a, t) = \cos\,(be^{ct}), 
\ \ {\rm with \ } x_{\rm ct}(a, 0) = \cos\,b = a,
\label{eqsens}
\end{equation}
where  $c > 0$. The motion described by 
Eq. (\ref{eqsens}) is bounded for all finite $t$;  
however, it shows sensitive dependence on initial conditions. 
The ratio of 
the distance $d_{\rm ct}(t)$ between two trajectories emanating from two 
infinitesimally close-by points $(a+\Delta)$ and $a$ is given by 
\begin{equation}
\nonumber
d_{\rm ct}(t) = \left\vert\frac{x_{\rm ct}(a+\Delta, t) - 
x_{\rm ct}(a, t)}{\Delta}\right\vert = 
\left\vert\frac{\partial x_{\rm ct}(a, t)}{\partial a}\right\vert
\end{equation}
to the first order in $\Delta$. It is easily seen that 
\begin{equation}
\nonumber
d_{\rm ct}(t) = \frac{1}{\sqrt{1-a^2}}
\left\vert\sin{(be^{ct})}\right\vert e^{ct},
\end{equation}
so that the Lyapunov exponent 
\begin{equation}
\nonumber
\lim_{t \to \infty} \frac{1}{t}\log{(d_{\rm ct}(t))} = c
\end{equation}
is positive, signaling chaos. 
Using Eq. (\ref{eqnonlocal2}), the corresponding 
expression for the distance in the discrete-time problem is
\begin{equation}
\nonumber
d_{\rm dt}(n) = 
\left\vert\frac{\partial}{\partial a}x_{\rm dt}(n)\right\vert = 
\frac{1}{\sqrt{1-a^2}} \frac{1}{(n-1)!} 
\left\vert\int_0^{\infty} du \,u^{n-1} 
\,e^{-u} \,e^{c\tau u}\,\sin\,(b e^{c\tau u})\right\vert. 
\end{equation}
Integrating once by parts, we get
\begin{equation}
\nonumber
d_{\rm dt}(n) = 
\frac{1}{\sqrt{1-a^2}} \frac{1}{bc\tau}\frac{1}{(n-1)!} 
\left\vert\int_0^{\infty} du \,
\cos\,(b e^{c\tau u}) \left[(n-1) u^{n-2} -u^{n-1}\right] \,
e^{-u}\right\vert.
\end{equation}
Using the fact that $\left\vert\int_0^{\infty} du \,\cos\,
(b e^{c\tau u}) 
f(u)\right\vert \le \int_0^{\infty} du \,\left\vert f(u)\right\vert$, 
we get  
\begin{equation}
\nonumber
d_{\rm dt}(n) \le  \frac{2}{bc\tau}\frac{1}{\sqrt{1-a^2}},
\end{equation}
which is bounded for all times. Thus the Lyapunov exponent 
for the discrete time 
evolution vanishes for this example.

\medskip
In order to get a feel for the differences between discrete and continuous 
time mechanics, let us consider a few more examples. 
If the equations of motion are 
such that $x_{\rm ct}(t) \sim t^{\alpha}$ (with $\alpha$ $>$ $-1$), then the 
corresponding discrete time evolution is given by $x_{\rm dt}(n)  
\sim \tau^{\alpha} \Gamma(n+\alpha)/(n-1)!$ 
which goes as $(n\tau)^{\alpha}$ for 
$\,n >> \alpha$. That is,  a system that shows power law evolution in 
continuous time shows similar behaviour in discrete-time mechanics. If 
$x_{\rm ct}(t) = a \,e^{bt} \,(b> 0)$ in continuous 
time, then the corresponding evolution in discrete time is given by 
$x_{\rm dt}(n) =  a \,e^{c\tau n}$ with 
$c = -(1/\tau)\,\log\,(1-b\tau) > b$. 
This implies that for systems that show unstable behaviour in continuous time, 
the instability is enhanced in the discrete-time context. 

\section{Conclusions}
To conclude, we have shown that the backward difference scheme preserves the 
non-negative character of the phase space density of a classical 
Hamiltonian system. The time evolution operator is a bounded 
operator for all times. We have shown that discretization of 
time leads to stochasticity. Irrespective of 
the number of degrees of freedom, one function of the phase space 
variables becomes stochastic. This variable has a unique probability 
density, which turns out to be the gamma density. 
The discrete-time evolution by a 
finite amount in the forward direction depends on the entire forward-time 
history of the continuous-time evolution. In this sense, the discrete-time 
evolution is nonlocal in time. The formula that relates 
$\rho_{\rm dt}(\vec{x}\,,\,\vec{p}\,, n)$ to $\rho_{\rm
  ct}(\vec{x}\,,\,\vec{p}\,,\, t)$ 
is suggestive of a random walk in an internal time. The motion 
of even free particles becomes correlated motion, in our formalism.  

\medskip
The same formalism is amenable to a quantum mechanical treatment 
via density 
matrices (equivalently, via the Wigner distribution functions). The elements 
of the density matrix that connect degenerate states are
time-invariant, whereas 
the ones that connect non-degenerate states decay exponentially. Thus 
discretization of time leads to energy super-selection 
and decoherence in quantum mechanics. It is also interesting to note
that 
an arrow of time emerges in the 
present framework. Finally, our approach does not seem to permit a 
consistent description of quantum mechanics via the discrete-time 
Schr\"odinger equation. 

\medskip
Some of the consequences of the discretization of time 
that we have pointed out in the foregoing are intriguing, and warrant 
further investigation to elucidate their meaning and implications. The 
complexities that may arise when the Liouville operator is not sufficiently 
`{\it good}' should also be explored. 

\medskip
To conclude, classical stochasticity, quantum decoherence and an arrow of 
time emerge automatically in the present version of discrete-time 
mechanics. All these features disappear and we recover the usual classical 
and quantum mechanics in the limit $\tau \to 0$.

\appendix
\section{Calculation of $g_n(\xi_1)$}
We now proceed to calculate $g_n(\xi_1)$ given by the relation 
\begin{equation}
\nonumber
g_n(\xi_1) = \left(\frac{1 - \alpha\tau\,\partial/\partial\xi_1}
{1 + \beta\tau\,\partial/\partial\xi_1}\right)^n \,\delta\,
\big(\xi_1 - \xi_1(0)\big).
\label{eqgnx1}
\end{equation}
The Fourier transform $\tilde{g}_n(k)$ of $g_n(\xi_1)$ is given by
\begin{equation}
\nonumber
\tilde{g}_n(k) = \int_{-\infty}^{\infty} \!d\xi_1 \,
e^{ik\xi_1} \,g_n(\xi_1) 
= \left(\frac{1+ i\alpha k}{1 - i\beta k}\right)^n e^{ik\xi_1(0)}.
\end{equation}
Rewriting $(1+ i\alpha k)/(1 - i\beta k)$ as 
$\beta^{-1}[(1 - i\beta\tau k)^{-1} - \alpha]$ 
and using the binomial theorem, we get
\begin{equation}
\tilde{g}_n(k) = \left(\frac{1}{\beta}\right)^n \sum_{l=0}^n 
{n \choose j} \,(-\alpha)^{n-j}
\left(\frac{1}{1 - i\beta\tau k}\right)^j \,e^{ik\xi_1(0)}.
\label{eqgnk}
\end{equation}
Taking the inverse Fourier transform of Eq. (\ref{eqgnk}), we get
\begin{equation}
g_n(\xi_1) = \left(-\frac{\alpha}{\beta}\right)^n \delta\,
\big(\xi_1 - \xi_1(0)\big) + 
\left(\frac{1}{\beta}\right)^n \sum_{j=1}^n 
{n \choose j}\, (-\alpha)^{n-j}\, 
h_j(\xi_1),
\label{eqgnxi11}
\end{equation}
where $h_j(\xi_1)$ is given by the relation
\begin{equation}
\nonumber
h_j(\xi_1) = \begin{cases}\frac{(\beta\tau)^{-j}}{(j-1)!} 
\left(\xi_1 - \xi_1(0)\right)^{j-1} 
\exp{[-\frac{(\xi_i - \xi_1(0))}{\beta\tau}]}, 
\ \ {\rm for} \ \xi_i > \xi_1(0) \cr \ 0, \quad\quad\,\,\,
\quad\quad\quad\quad\quad\quad\quad\quad\quad\quad\quad\quad\quad   \ \ 
{\rm for} \ \xi_i < \xi_1(0).
\end{cases}
\end{equation}
In Eq. (\ref{eqgnxi11}), the first term is singular, while the 
second is the sum of 
a finite number of regular functions. 
The first term is negative for odd $n$. It 
then follows that $\alpha$ has to be zero 
for the probability density 
$g_n(\xi_1)$ to be positive for all $n$. 
In that case $g_n(\xi_1) = h_n(\xi_1)$.

\section{Time evolution of $\rho_{\rm dt}$}

For forward evolution, i.e., $n >0$, define the generating function
\begin{equation}
\nonumber
G(\vec{x}\,,\,\vec{p}\,,\,z) = \sum_{n=0}^{\infty}z^n \,
\rho_{\rm dt}(\vec{x}\,,\,\vec{p}\,,\,n) = 
\sum_{n=0}^{\infty}z^n\left(\frac{1 + \alpha\tau L}{1 - \beta\tau L}\right)^n 
\rho_{\rm dt}(\vec{x}\,,\,\vec{p}\,,\,0),
\label{eqgenerate1}
\end{equation}
to get
\begin{equation}
\nonumber
G(\vec{x}\,,\,\vec{p}\,,\,z) = 
\left(\frac{1 - \beta\tau L}{1-z -(\beta + \alpha z)\tau L}
\right) \rho_{\rm dt}(\vec{x}\,,\,\vec{p}\,,\,0).
\end{equation}
This expression can be further simplified to
\begin{equation}
\nonumber
G(\vec{x}\,,\,\vec{p}\,,\,z) = \left(\frac{\beta}{\beta + \alpha z}\right) 
\rho_{\rm dt}(\vec{x}\,,\,\vec{p}\,,\,0) 
+ \left(\frac{z}{\beta + \alpha z}\right)
\left(\frac{1}{1-z -(\beta + \alpha z)\tau L}\right) 
\rho_{\rm dt}(\vec{x}\,,\,\vec{p}\,,\,0).
\end{equation}
Now use the integral representation
\begin{eqnarray}
\nonumber
\frac{1}{1-z -(\beta + \alpha z)\tau L} &=&
\int_0^{\infty}\! du \,\exp\,
{\left\{-u \,[1-z -(\beta + \alpha z)\tau L]\right\}}\\
\nonumber
&=& \sum_{m=0}^{\infty}\frac{z^m}{m!} 
\left(1 + \alpha\tau L\right)^m \int_0^{\infty}\!du 
\,u^m\,e^{-u} \,e^{\beta\tau L},
\end{eqnarray}
to get
\begin{eqnarray}
\nonumber
G(x,p,z) &=&  \left(\frac{\beta}{\beta + \alpha z}\right)
\rho_{\rm dt}(\vec{x}\,,\,\vec{p}\,,\,0) +  
\left(\frac{z}{\beta + \alpha z}\right)
\sum_{m=0}^{\infty}\frac{z^m}{m!} 
(1 + \alpha\tau L)^m \nonumber \\
&& \times \int_{0}^{\infty} du \, u^m \,e^{-u} 
\,\rho_{\rm ct}(\vec{x}\,,\,\vec{p}\,,\,\beta\tau u),
\end{eqnarray}
where $\rho_{\rm ct}(\vec{x}\,,\,\vec{p}\,,\,t)$ 
is the continuous-time phase space 
density given by Eq. (\ref{eqformal}). 
Collecting together the coefficients of $z^n$  yields
\begin{eqnarray}
\nonumber
\rho_{\rm dt}(\vec{x}\,,\,\vec{p}\,,\,n) &=& 
\left(-\frac{\alpha}{\beta}\right)^n 
\rho_{\rm dt}(\vec{x}\,,\,\vec{p}\,,\,0) + 
\frac{1}{\beta} \sum_{j=0}^{n-1} \left(-\frac{\alpha}{\beta}\right)^{n-1-j} 
\frac{(1 + \alpha\tau L)^j}{j!}\nonumber \\
&&\times  \int_{0}^{\infty} \!du \, u^j \,e^{-u} 
\,\rho_{\rm ct}(\vec{x}\,,\,\vec{p}\,,\,\beta\tau u).
\label{eqgenerate}
\end{eqnarray}
The first term on the right-hand side 
in Eq. (\ref{eqgenerate}) is singular, and negative for 
odd $n$. The second is a sum of finite-order derivatives of a regular  
function. Therefore, in order for 
$\rho_{\rm dt}(\vec{x}\,,\,\vec{p}\,,\,n)$ to be non-negative, $\alpha$ must
vanish. Thus, both the methods referred to in the main text lead to  
the same answer. However, the second method yields the representation
\begin{equation}
\nonumber
\rho_{\rm dt}(\vec{x}\,,\,\vec{p}\,,\,n) = \frac{1}{(n-1)!} 
\int_{0}^{\infty}\! du \, e^{-u} \,u^{n-1} 
\,\rho_{\rm ct}(\vec{x}\,,\,\vec{p}\,,\,\tau u)
\end{equation}
for $\rho_{\rm dt}(\vec{x}\,,\,\vec{p}\,,\,n)$ in terms of 
$\rho_{\rm ct}(\vec{x}\,,\,\vec{p}\,,\,t)$.
\end{document}